\documentclass[
preprint,
superscriptaddress,
showpacs,
amsmath,amssymb,
prl,
]{revtex4-1}

\usepackage{graphicx}
\usepackage{dcolumn}
\usepackage{bm}
\usepackage[breaklinks,colorlinks = true,linkcolor = red,urlcolor=blue,citecolor=blue]{hyperref}
\usepackage{multirow}
\usepackage{array}
\usepackage{booktabs}
\usepackage{ctable}
\usepackage{upgreek}
\usepackage{epsfig}
\usepackage{mathrsfs}
\usepackage{amssymb}
\usepackage{amsbsy}
\usepackage{color}
\usepackage{cancel}
\usepackage{pifont}
\usepackage{marginnote}
\usepackage{float}
\usepackage{verbatim}
\usepackage{subfigure}

\begin{document}
\title{Carrier dynamics in ultrathin gold nanowires: Role of Auger processes}

\author{Gyan Prakash} \affiliation{Department of Physics and Center for Ultrafast Laser Applications, Indian Institute of Science, Bangalore 560012, India}

\author{Subhajit Kundu} \affiliation{Materials Research Centre, Indian Institute of Science, Bangalore 560012, India}

\author{Ahin Roy} \affiliation{Materials Research Centre, Indian Institute of Science, Bangalore 560012, India}

\author{Abhishek K. Singh} \affiliation{Materials Research Centre, Indian Institute of Science, Bangalore 560012, India}

\author{N. Ravishankar} \affiliation{Materials Research Centre, Indian Institute of Science, Bangalore 560012, India}

\author{A. K. Sood}
\affiliation{Department of Physics and Center for Ultrafast Laser Applications, Indian Institute of Science, Bangalore 560012, India}

\date{\today}
\begin{abstract}  
Carrier dynamics in metallic nanostructures is strongly influenced by their confining dimensions. Gold nanoparticles of size $\sim 2$ nm lie at the boundary separating metallic and non-metallic behavior. In this work, we report carrier dynamics in high aspect ratio ultrathin gold nanowires (Au-UNWs) of average diameter $\sim$2 nm using  pump (3.1 eV) and coherent white light continuum as probe in the spectral range of 1.15 eV to 2.75 eV. The transient carrier dynamics in Au-UNWs under extreme excitation regime is slower than predicted by the often used two temperature model. We identify Auger-assisted carrier heating process which slows down the hot carrier cooling dynamics. The rate equation model fitted to the data yields an estimate of Auger coefficient for gold nanostructures. 
\end{abstract}
\maketitle
\section{\label{sec:level1}  Introduction}
Motivated by both fundamental and interesting technological applications, the evolution of electron dynamics with size in metal nanosystems  has been  studied using ultrafast time-resolved optical techniques to understand the impact of size on electron-electron (e-e) thermalization and electron-phonon (e-ph) cooling dynamics in  metal nanoparticles (MNs) of gold (Au) and silver (Ag) \cite{link1999electron,minutella2017excitation,higaki2018sharp,zhou2016evolution,voisin2004ultrafast,arbouet2003electron,voisin2000size}. These studies have addressed the confinement and surface effects on the carrier dynamics of these MNs.  The carrier dynamics in MNs of size $>$ 10 nm was found to be similar to the bulk system \cite{link1999electron,minutella2017excitation},  revealing a strong size dependent dynamics  for MNs $<$ 10 nm. Excluding the gold nanoparticles in the size regime $<$ 2 nm, where a transition from metallic to molecular behavior is observed due to extreme confinement \cite{zhou2016evolution,higaki2018sharp}, the MNs of Ag and Au showed rapid thermalization and a faster electron-phonon mediated cooling dynamics with decreasing size from 10 nm to 2 nm \cite{voisin2004ultrafast,arbouet2003electron,voisin2000size}.  The observations were ascribed to the conduction electron spillout and d-band localization due to the surface, leading to reduced screening of  e-e and e-ion interactions \cite{voisin2004ultrafast,arbouet2003electron,voisin2000size,ekardt1984work}. Electron dynamics in ultrathin metal nanowires  which confine the conduction electrons only in two dimensions has not been explored yet.

Recent chemical synthesis methods have been successfully utilized to grow ultrathin gold nanowires (Au-UNWs) with diameter $\sim$ 2 nm  and length on the $\mu$m scale \cite{halder2007ultrafine}.  The atomically precise Au-UNWs  having high mechanical flexibility with aspect ratio $>$ 1000 have been used in fabricating flexible  transparent conductive metal grids \cite{maurer2016templated}, fast response pressure sensors \cite{gong2014wearable},  mechanical energy storage \cite{xu2010mechanical} and interconnects in nano-devices \cite{lu2010nanoelectronics}.  The dominant edge sites on their surfaces have been explored for active and  selective  electrochemical reductions \cite{zhu2014active} with  applications  in the fields of catalysis \cite{jiang2018highly}. Au-UNWs till now have been  studied for their  mechanical strength \cite{chen2013mechanically}, wrinkling of atomic planes \cite{roy2014wrinkling}, tunable electronic property \cite{chandni2013tunability}   and observing broad band longitudinal mode of the surface plasmon resonance in the IR region \cite{takahata2014surface}. This work focuses on  studying photoexcited carrier dynamics in Au-UNWs with a view to understand the effect of two dimensional confinement on the carrier dynamics of gold.

   Coulomb e-e interaction is  enhanced as the size of metals or semiconductors are reduced from bulk to nanoscale dimensions \cite{klimov2000quantization,voisin2004ultrafast,arbouet2003electron,voisin2000size,ekardt1984work}. The efficiency of multiparticle interactions like Auger-mediated interactions are, therefore, also enhanced with reduced dimensions \cite{klimov2000quantization}. Here, using ultraviolet (UV) pump  of photon energy 3.1 eV, greater than the interband transition threshold (2.3 eV) of gold,  we photoexcite the carriers in chemically synthesized Au-UNWs (diameter $\sim$ 2 nm and length $\sim$ 1 $\mu$m) and monitor the subsequent evolution of the transient dynamics of the  excited carriers in a broad spectral range using  coherent white light probe.   We systematically probe the dynamics  by varying the pump intensity from weak to strong excitation regime and show that the mechanism of carrier relaxation through electron-phonon  observed so far in larger size Au nanostructures is not sufficient in the Au-UNWs. We show that the reduced screening of  e-e interaction due to spilling of conduction band  and localization of core (d band) electrons at the surface promotes Auger heating. Explicit signature of this is clearly seen in the transient kinetic traces exhibiting slower cooling dynamics than predicted by the two temperature model (TTM). The transient dynamics is modeled  using rate equations including both first-order and Auger mediated processes which capture the full relaxation dynamics, giving an estimate of the Auger-coefficient.  

\section{\label{sec:level2}  Experimental and Computational Details}
\begin{figure*}[ht]
\includegraphics[width=1\linewidth]{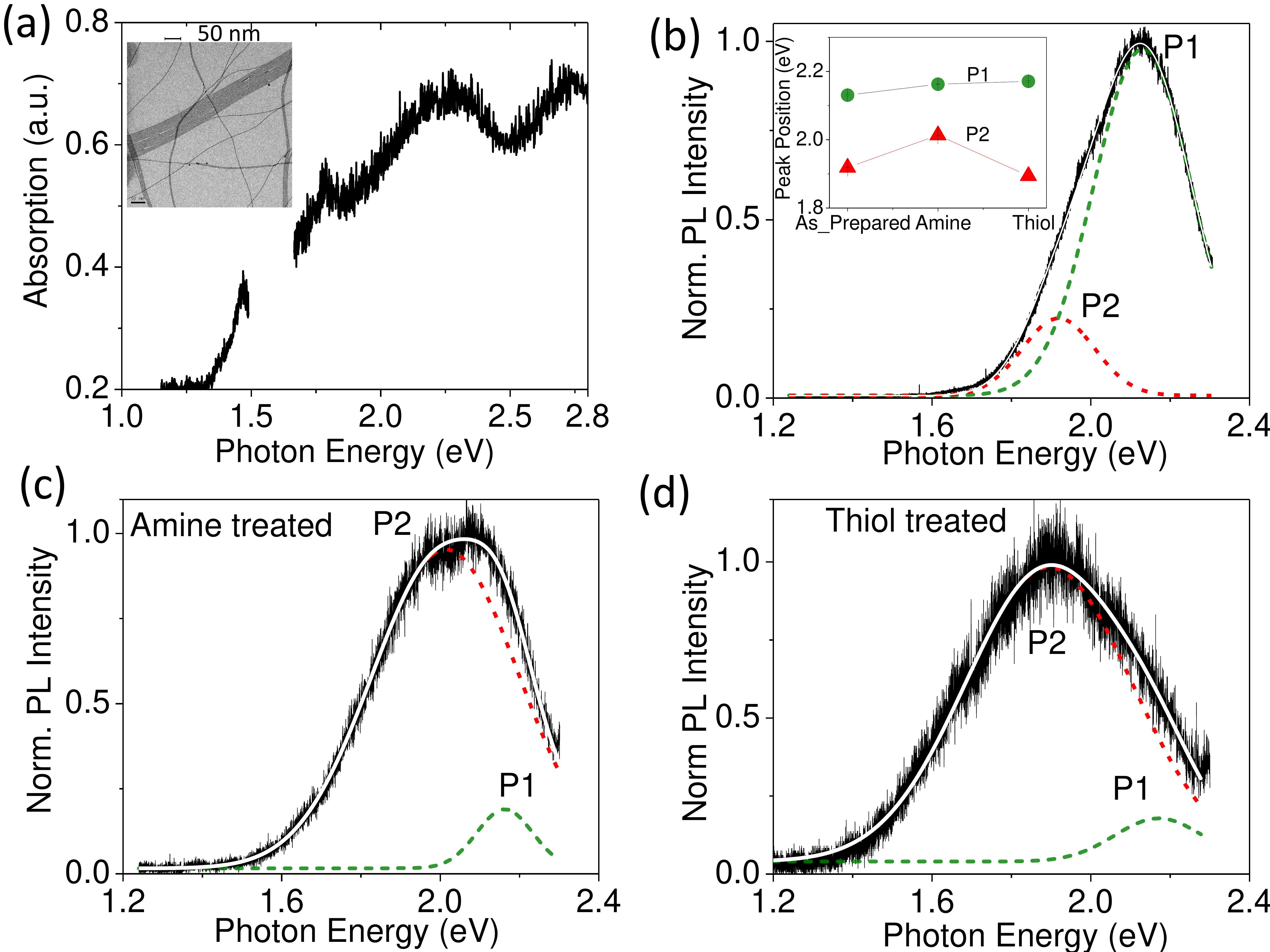}
\caption{(a) Steady state absorption of Au-UNWs measured with unexcited coherent white light continuum. The missing data points around 1.55 eV are due to the removal of the residual scattering of the white-light continuum generating beam. Inset shows the TEM image of aligned and individual long Au-UNWs. PL emission emission spectrum  of the  (b) as prepared Au-UNWs sample and samples after treatment with (c) n-Butyl amine and (d) 1,4-Butanedithiol on Si substrate. Solid lines are fit to emission spectrum with a multi-peak Gaussian lineshape function. Individual Gaussians contributing to the total fit are shown by dahsed lines. Inset in (b) shows the peak positions obtained from the fit in (b-d).} 
\label{fig:SEM_Abs}
\end{figure*}

Ultrathin Au nanowires  were synthesized by our published method \cite{halder2007ultrafine}. In brief, 6 mg of  HAuCl$_4$ was dissolved in 5 ml of n-hexane by complexation with 200 $\mu$l of oleylamine. Subsequently, 300 $\mu$l of tri-isopropylsilane was added to it. The solution was aged at room temperature for $\sim$1 day.  
10 ml of absolute ethanol was added to 5 ml of the as-prepared Au nanowire solution and shaken. The mixture was centrifuged at $\sim$2500 rpm for $\sim$2 mins. The precipitate obtained was re-dispersed in $\sim$2 ml of n-hexane to obtain concentrated Au nanowire solution.  To confirm the morphology, the same solution was drop-casted onto electron transparent membrane for TEM imaging. It showed formation of ultrathin Au nanowires along with trace amounts of Au nanoparticles. The Au nanowire solution was centrifuged at 4000 rpm without addition of ethanol to reduce the presence of nanoparticles to negligible amounts as confirmed by TEM imaging. The precipitate obtained was redispersed in n-hexane and drop-casted  on  standard glass of 1 mm thickness and Si substrates of thickness $\sim$0.5 mm cut into pieces of size $10\times 10$ mm   for ultrafast differential transmission and PL measurements, respectively. 

Ultrafast transient differential transmission experiments were done using excitation pulses of 400 nm (3.1 eV)  obtained by frequency doubling the $\sim$80 fs  pulses of central frequency 800 nm (1.55 eV) from the output of a Ti:Saphire amplifier (Spitfire from M/s Spectra Physics Inc. operating at 1 KHz repetition rate) with a BBO crystal. The white light continuum used as probe in the transient differential transmission spectroscopy set up (`ExciPro' from M/s CDP systems Corporation) was obtained by focusing a small fraction of 800 nm (1.55 eV) pulses on a sapphire crystal. A spectrometer based on diffraction grating and two channel linear-array of photodiodes was used to record pump-induced transmission changes in the spectral range of 1.15 eV to 2.75 eV. The 400 nm pump beam was modulated at 250 Hz using a mechanical chopper for phase locked detection.

The dielectric function of Au-UNWs was calculated using first-principles density functional theory (DFT). The details of computation are as follows. The Au nanowire, oriented along [111] direction with a $\sim$2 nm diameter, was cleaved from bulk Au. The wire was terminated with six low-index \{110\} surfaces, leading to hexagonal-prism morphology, as shown in Fig. 2 (a). The cell length along the nanowire axis and the atomic positions in the unit cell have been optimized without any symmetry constraint. More than 10~\AA $ $ vacuum space was included in the transverse direction to avoid interactions among the periodic replicas of the systems. The calculations were performed using first principles density functional theory (DFT), wherein the ionic cores are described by all-electron projector augmented wave potentials~\cite{PAW1,PAW2} and the Perdew-Burke-Ernzerhof ~\cite{PBE} generalized gradient approximation to the electronic exchange and correlation as implemented in the Vienna Ab Initio Simulation Package (VASP) \cite{PAW1,PAW2,Kresse}. Geometries were considered converged when the component of interatomic forces were less than 0.005 eV/~\AA. The one-dimensional Brilluoin zone integration was performed with a well-converged Monkhorst-Pack ~\cite{pack} with \textbf{k}-grid of 1$\times$1$\times$5. The favourable binding site of the CH$_3$S and CH$_3$NH$_2$ were found by calculating the binding energies at different inequivalent sites of the nanowire.  Once the electronic ground-state is determined, the frequency ($\omega$) dependent imaginary part of the dielectric function was determined by a summation over empty states:

\begin{equation}
\epsilon_{\alpha \beta}^2 (\omega)= \frac{4\pi{^2}e{^2}}{\Omega} \lim_{q \to 0}\frac{1}{q^2} \sum_{c,v,k} 2\omega_{k} \delta(\epsilon_{ck}-\epsilon_{vk}-\omega) \times \langle {u_{ck+c_{\alpha q}} | u_{vk}} \rangle \langle {u_{ck+c_{\beta q}} | u_{vk}} \rangle
\end{equation} 


where, $\alpha$ and $\beta$ are different cartesian directions, indices $\it c$ and $ \it v$ refer to conduction and valence band states respectively, and $ u_{c\mathbf{k}}$ is the cell periodic part of the orbitals at the k-point $\bf k$.

\section{\label{sec:level3}  Results and Discussion}
\begin{figure*}
\includegraphics[width=1\linewidth]{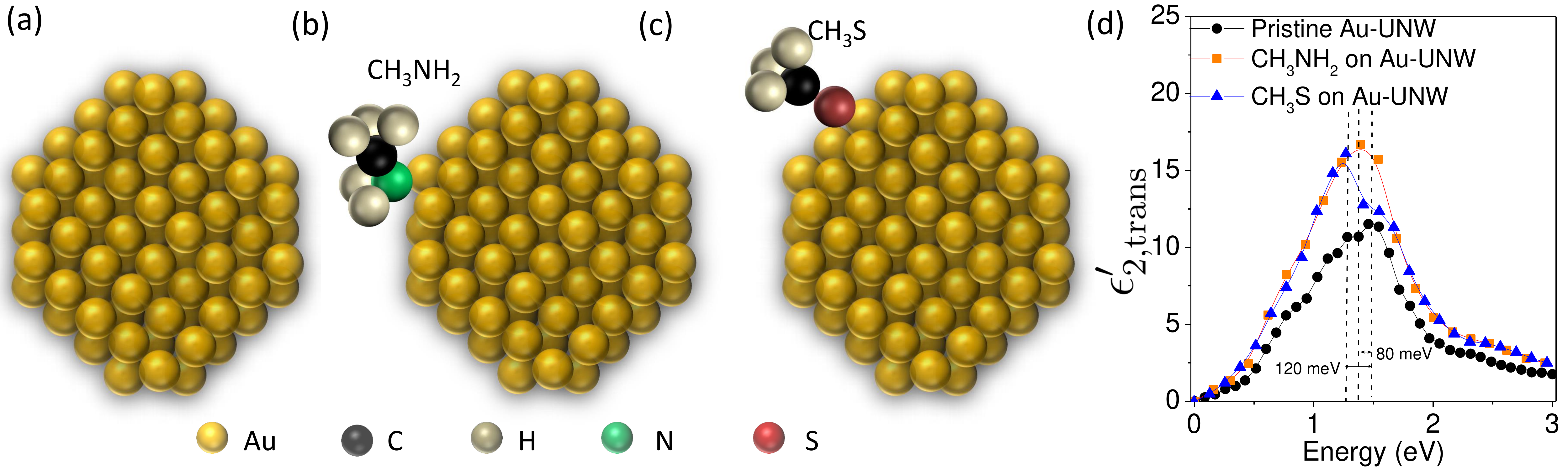}
\caption{Relaxed structures of (a) pristine  (b) amine-adsorbed and (c) thiol-adsorbed gold nanowires of $\sim 2$ nm diameter.  The number of atoms shown in the figure are representative only. Actual simulations were done with 169 atoms in a unit cell. (d) Imaginary part of transverse dielectric function of pristine and adsorbed Au nanowire.} 
\label{fig:epsilon_x_imag}
\end{figure*}

Inset of Fig. \ref{fig:SEM_Abs} (a) shows the TEM image of long Au-UNWs  and Fig. \ref{fig:SEM_Abs} (a) displays  steady state absorption taken after drop-casting on a  glass substrate. The steady state absorption  was obtained using the white light continuum of transient differential transmission spectroscopy set up by keeping the pump beam blocked. The absorption spectrum shows a peak at $\sim$2.3 eV. We attribute the origin of the absorption peak to the transverse plasmon resonance \cite{tcherniak2011one,mohamed2000thelightning,varnavski2003relative} of Au-UNWs. We also record the PL emission spectrum of the Au-UNWs to understand  the  origin of observed absorption peak due to two possibilities: transverse plasmon resonance and absorption by the amine ligands attached at the surface of the Au-UNWs during the synthesis process \cite{liu2016luminescent}.  Fig.  \ref{fig:SEM_Abs} (b)  shows the PL emission spectrum of Au-UNWs on silicon (Si) substrate recorded with 532 nm (2.33 eV) excitation wavelength. The PL emission can be decomposed into two bands with their peak positions at 2.13 eV and 1.91 eV, marked as P1  and P2, respectively.  Solid line  in Fig. \ref{fig:SEM_Abs} (b) represents the fit to the PL spectrum with the sum of two Gaussians \cite{lei2001preparation,link2002visible} and the dashed lines show the individual Gaussians.  To understand the origin of  P1 and P2 PL emission bands, we changed  surface functionalization of the nanowires by treating with n-Butyl amine (amine) and 1,4-Butanedithiol (thiol). The treatment of the Au-UNWs with n-Butyl amine and 1,4-Butanedithiol increases the coverage of Au-N and Au-S centers on the surface of the Au-UNWs, respectively.   The PL spectrum after treating the sample with n-Butyl amine and 1,4-Butanedithiol along with the fitting  are shown in Fig. \ref{fig:SEM_Abs} (c) and (d), respectively. The intensity of the P1 band is very low as compared to P2 band in the n-Butyl amine as well as in 1,4- Butanedithiol treated Au-UNWs.  As shown in inset of Fig. \ref{fig:SEM_Abs}(b)  the peak position of P1 band shows a  slight blue-shift of 0.02 eV (0.03 eV) for amine (thiol) treated nanowires, in comparison to a blue-shift of 0.09 eV for amine treated and a red-shift of 0.03 eV of P2 after thiol treatment. As the emission  shows only  a very small shift in its peak position after additional surface treatments and its intensity decreases on surface treatment, we attribute P1 band to the radiative decay of transverse plasmons \cite{imura2009properties,tcherniak2011one,liu2017relaxation}. A slight blue shift of the P1 band arises from the changed dielectric constant due to surface functionalization.  The emission corresponding to the P2 band is attributed to the emission from Au-N and Au-S defect centers at the surface of Au-UNWs \cite{liu2016luminescent}. A small intensity of the P1 emission after surface treatments is because of the availability of the large Au-N or Au-S defect centers through which the transverse plasmon decays nonradiatively  suppressing the emission from the radiative decay \cite{liu2016luminescent}.  We note that  the emission peak P1 is red-shifted with respect to the absorption peak by $\sim$170 meV, similar to that has been observed in gold nano particles of diameter $\sim$ 2 nm \cite{dulkeith2004plasmon} and is attributed to Stokes shift.

Origin of the observed emission peaks due to transverse plasmon (P1) and absorbed amine or thiol ligands (P2) on the surface of the Au-UNWs were also confirmed by the first-principles density functional theory (DFT) calculations. In order to probe the effect of adsorption on the optical property of Au-UNWs, simplified alkyl chains of methylamine (CH$_3$NH$_2$) and methanethiolate (CH$_3$S) were adsorbed on the energetically most favorable site \cite{roy2014semiconductor}. CH$_3$NH$_2$ binds at an edge site whereas CH$_3$S prefers a two-fold binding at an FCC surface site  as shown in Fig. \ref{fig:epsilon_x_imag} (a-c) \cite{roy2014semiconductor}. Using Perdew-Burke-Ernzerhof generalized gradient approximation to the electronic exchange and  VASP for implementing the electronic correlations, we calculate the transverse dielectric function of the Au-UNW ($\epsilon_{\textrm{trans}} (\omega)=\epsilon_{1,\textrm{trans}}(\omega)+i\epsilon_{2,\textrm{trans}}^{\prime}(\omega)$), where $\epsilon_{1,\textrm{trans}}(\omega)$ and $\epsilon_{2,\textrm{trans}}^{\prime}(\omega)$ are the real and imaginary parts of the complex dielectric function. A plot of the imaginary part of transverse dielectric function $\epsilon_{2,\textrm{trans}}^{\prime}(\omega)$ for pristine, CH$_3$NH$_2$ adsorbed and CH$_3$S adsorbed Au-UNWs is shown in Fig. \ref{fig:epsilon_x_imag}(d). The observed peak in $\epsilon_{2,\textrm{trans}}^{\prime}(\omega)$ for pristine Au-UNW is attributed to the transverse plasmon resonance.  As DFT underestimates the band gap \cite{perdew1985density}, the transverse plasmon resonance peak in the simulated $\epsilon_{2,\textrm{trans}}^{\prime}(\omega)$ for pristine Au-UNW is $\sim$0.6 eV lower than that observed experimentally.  Nevertheless, the simulated $\epsilon_{2,\textrm{trans}}^{\prime}(\omega)$ agrees well with the key features observed in the emission spectrum. Similar to the emission spectrum of thiol treated Au-UNW a peak in $\epsilon_{2,\textrm{trans}}^{\prime}(\omega)$ due to the adsorbed molecules of CH$_3$S on Au-UNWs is clearly seen at a red-shifted position with respect to the transverse absorption peak. Moreover, the red-shift of 120 meV is also qualitatively similar to the experimentally observed red-shift of 280 meV in thiol treated Au-UNW. The peak due to adsorption of CH$_3$NH$_2$ is, however, not well separated in the simulated $\epsilon_{2,\textrm{trans}}^{\prime}(\omega)$ for the CH$_3$NH$_2$ adsorbed Au-UNW. The well separated peaks in CH$_3$S adsorbed Au-UNW are due to aurophlic nature of CH$_3$S which leads to stronger hybridization compared to that of CH$_3$NH$_2$.

\begin{figure}
\includegraphics[width=1\linewidth]{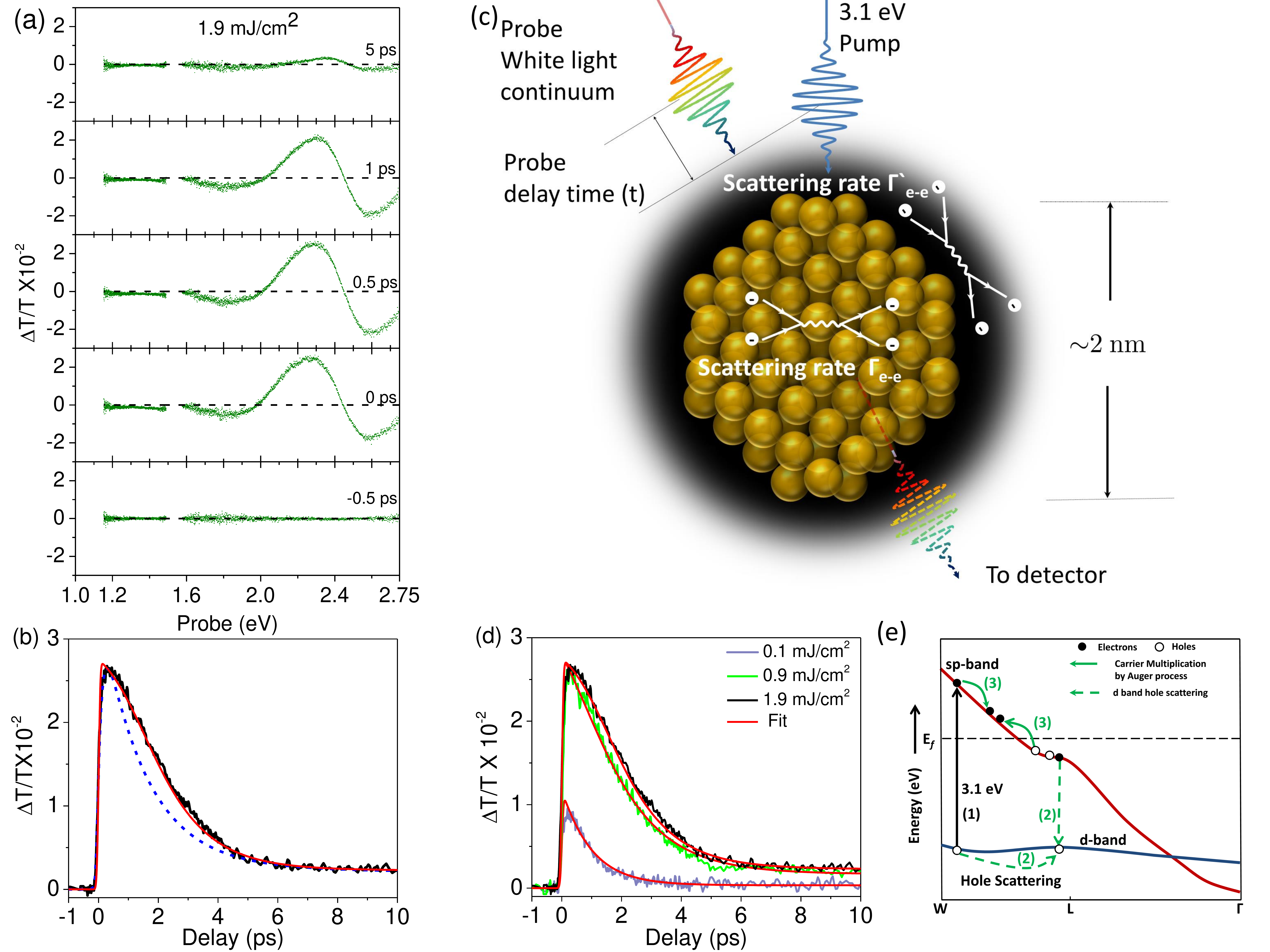}
\caption{(a) Pump induced change in probe transmission at -0.5 ps, 0 ps, 0.5 ps, 1 ps and 5 ps delay for pump fluence 1.9 mJ/cm$^2$. Dashed horizontal lines mark $\Delta T/T=0$ for each plot.  (b) Kinetic trace of the data shown in (a) at probe energy 2.3 eV. The dashed line is fit to the data using TTM.  Solid line in red is fit to the data considering the first-order decay together with  free carrier three-body Auger heating process. Inset shows the same data on a logarithmic scale. (c) Schematic illustration showing the cross sectional view of Au-UNW. The conduction electron spillout regions close to the surface of Au-UNW are shown in grey. Scattering rate of electrons close to the surface $\Gamma_{e-e}^\prime$(surface)$>\Gamma_{e-e}$(inner region).  (d) Kinetic traces  at three fluences (0.06, 0.30 and 1.50 mJ/cm$^2$) at probe energy 2.3 eV. Solid line in red is fit to the data considering the first-order decay together with  free carrier three-body Auger process. (e) Schematic illustrating (1) interband (d to sp band)  photoexcitation, (2) d band hole scattering and  (3) Auger heating process.  Band structure of gold adopted from the literature \cite{dulkeith2004plasmon}.} 
\label{fig:Abs_TA}
\end{figure}

We now present pump-induced differential transmission changes observed in Au-UNWs. Fig. \ref{fig:Abs_TA} (a) shows the pump induced transmission change $\Delta T/T$ observed in Au-UNWs after  -0.5 ps, 0 ps, 0.5 ps, 1 ps and 5 ps  delay time at a fixed pump fluence of 1.9 mJ/cm$^2$.   The differential transmission change exhibits a positive $\Delta T/T$ (bleaching) signal around 2.3 eV with two negative wings. The mechanism of the observed photobleach signal is a manifestation of two-step process involved immediately after the interband (5d to 6sp band) excitation of the photoexcited carriers by the pump: (i) the photoexcited electrons thermalize to form a population distribution of hot electrons in the conduction band at an elevated temperature $T_e>>T_L$ (initial lattice temperature) through electron-electron scattering  within  $\sim$ 100 fs to few hundreds of fs of thermalization time \cite{sun1994femtosecond,link1999electron,zhou2016evolution} (ii)  the nonequilibrium thermalized population of hot carriers excites the transverse plasmons \cite{hu2012plasmon} and simultaneously loses its energy to the lattice through electron-phonon interaction.  An equilibrium with the lattice at a slightly higher temperature than $T_L$ is then reached,  followed by cooling of the hot lattice on time scale of $\sim$ 100 ps. Thus the observed $\Delta T/T$ signal is due to the bleaching of the transverse plasmon absorption. The negative $\Delta T/T$ signals on both sides of the bleach signal are due to the broadening of the transverse plasmon band at increased electronic temperature \cite{yu2011excitation}.  
 Therefore, the transient differential transmission measurements probe the carrier dynamics  in Au-UNWs indirectly by monitoring the  transverse plasmon resonance bleach recovery \cite{link1999electron,harutyunyan2015anomalous,cai2018photoluminescence}.  

The cooling dynamics of the thermalized hot electrons has been explained by the well-established two-temperature model (TTM). The TTM describes the energy exchange rate between the electrons and phonons  by the two coupled heat equations $C_e {d T_e}/{dt}=-g \left(T_e-T_L\right)$  and $C_L {d T_L}/{dt}=g \left(T_e-T_L\right)$ where $C_e \left(C_L\right)$ and $g$ are heat capapcity of electrons (lattice) and electron-phonon coupling constant, respectively \cite{sun1994femtosecond}. Following the TTM we fit the kinetic traces at probe energy 2.3 eV, corresponding to the bleach maximum, with the following function convoluted with the Gaussian pulse response \cite{link1999electron,sun1994femtosecond}:
   \begin{equation}
  \frac{\Delta T}{T} = H(t)\left[A_{e,NTh} e^{-t/\tau_r'}+A_{e,Th}(1-e^{-t/\tau_r}) e^{-t/\tau_p}+A_L\right]
  \label{equtn_TTM}
  \end{equation}  

where H(t) is a Heaviside step function. $A_{e,NTh}$, and $A_{e,Th}$ are the initial transient amplitudes corresponding to non-thermalized distribution of electrons interacting with the phonon bath during the thermalization process and the thermalized  electrons, respectively. $A_L$ is
the offset amplitude due to slow cooling of the lattice. The time constant $\tau_r$ is the rise time associated with the thermalization of photoexcited carriers and $\tau_p$ is the time constant for the decay of the thermalized distribution due to electron-phonon interaction. Note $\tau_r'=\left(1/\tau_r+1/\tau_p\right)^{-1}$ is the decay time of non-thermalized population, taking into account simultaneous interaction of nonthermalized electrons with phonon bath \cite{sun1994femtosecond}.   Fig. \ref{fig:Abs_TA}(b) shows the kinetic trace at 2.3 eV together with the best fit (dashed line) to the experimental data using the fitting values of $\tau_r=148\pm 10$ fs, $\tau_p$=1.5 $\pm$0.1 ps, $A_{e,Th}=3.1\pm0.1 \times 10^{-2}$, $A_{e,NTh}=1.1 \pm0.1 \times 10^{-3}$ and $A_L=1.8\pm0.1 \times 10^{-3}$. 
The observed thermalization time of carriers $\sim 148$ fs, much faster than the well-characterized thermalization time of $\sim 500$ fs \cite{sun1994femtosecond} in the case of intraband excitations,  is due to the initial dynamics of electrons dominated by the formation of hot electrons via d-state Auger electron-hole recombination and is consistent with the reported thermalization time of $\sim 100$ fs  \cite{link1999electron,masia2012measurement} for inter-band excitation in gold. However, we clearly see that the  TTM is not able to capture the  decay dynamics of thermalized carriers. The experimentally  observed decay is slower than that predicted by the TTM upto the delay of 4.5 ps. 

To unravel the physical mechanism responsible for the slower dynamics in Au-UNWs we look at the possible factors which can affect the dynamics. Due to the kinematic restrictions on electrons in metal nanostructrues of size smaller than the mean free path, the electron-lattice interaction becomes significantly weaker \cite{bilotsky2006size,belotskii1992surface}.  The conduction electron wave functions at the surface are modified and the spatial distribution of  electrons are not homogeneous throughout the volume of Au-UNWs \cite{voisin2004ultrafast}. The spill-out of the conduction electrons and localization of the core electrons enhances the Coulombic electron-electron interactions close to the surface of Au-UNWs \cite{voisin2004ultrafast,robel2006exciton,gao2013transient}.  The reduced screening in low-dimensional systems thus promotes  many-body interaction Auger processes.  Auger-assisted carrier multiplication (CM) process (Auger heating) can slow down the hot carrier cooling \cite{brida2013ultrafast}. Fig. \ref{fig:Abs_TA} (c and e) illustrate the mechanism of free carrier three-body collision Auger process in Au-UNWs. 

In the following discussion, we model the carrier dynamics incorporating free carrier three-body collision (electron-electron-hole or hole-hole-electron) Auger process along with a first-order decay process accounting for electron-lattice energy exchange through electron-phonon interaction. The transient differential transmission kinetic is  proportional to non-equilibrium thermalized carrier density and the non-equilibrium phonons generated after the cooling of electrons through e-ph interaction. 

\begin{figure*}[ht]
\includegraphics[width=1\linewidth]{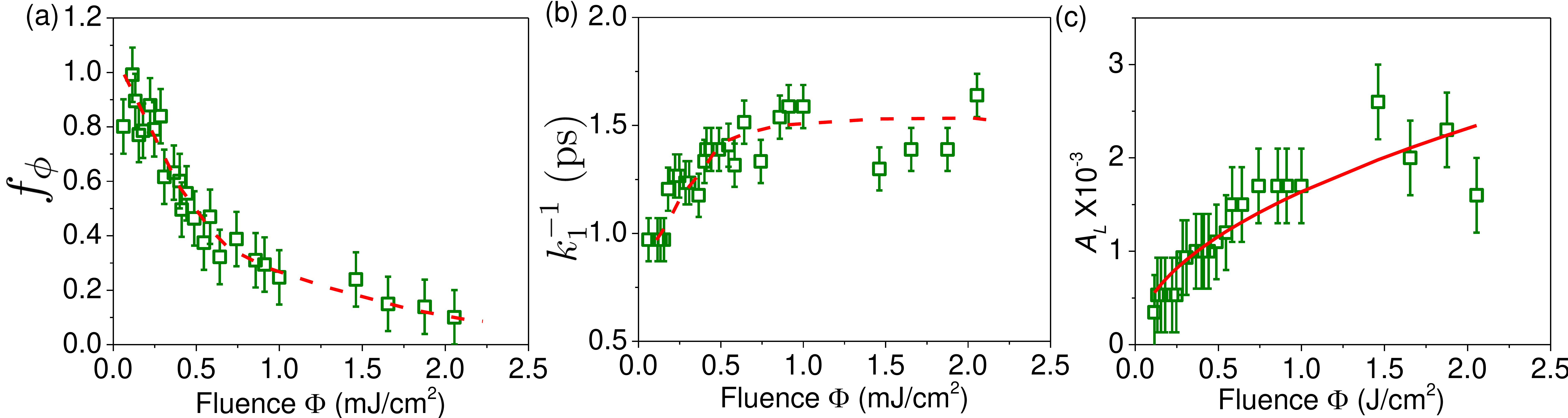}
\caption{Pump fluence dependence of (a) decay time constant ($\tau_p=k_1^{-1}$) (b) filling factor ($f_\phi$) and (c) $A_L$. The dashed lines in red through the data points are guide to eye. The solid in (c) is fit to the data points with a function $\sim \sqrt{\Phi}$}  
\label{fig:TA_param}
\end{figure*}

The rate equation for transient non-equilibrium carrier density can be written as \cite{trinh2015many}:
  \begin{equation}
  \frac{dN(t)}{dt}=f_\phi G(t)-k_1N(t)+k_AN(t)^3
  \label{equtn1}
  \end{equation}

where $N(t)$ is the transient free carrier density (electrons or holes) at time $t$, $k_1$ is first-order rate constant and $k_A$ is Auger coefficient (third-order rate constant).  $G$ and $f_\phi$ are the pump induced generation rate  and the filling factor of the free carriers, respectively.  The transient differential transmission kinetic traces at the photobleach maximum (2.3 eV) are then modeled as

  \begin{equation}
  \frac{\Delta T}{T} \propto \left[ N(t) + A_L e^{-t/\tau_L}\right]
  \label{equtn2}
  \end{equation}

where $A_L$ is the amplitude of the transmission change due to the hot lattice and $\tau_L$ is its cooling time constant $\sim$ 1 ns \cite{corkum1988thermal,sun1994femtosecond}.  As shown in Fig. \ref{fig:Abs_TA} (b)  the theoretical fit with equation (\ref{equtn2}) captures the  decay dynamics very well.  The kinetic traces at 2.3 eV for all pump fluences were fitted by keeping $k_A$ and $\tau_L$ fixed and varying $k_1$, $f_\phi$ and $A_L$ as a function of pump intensity. Fig. \ref{fig:Abs_TA} (d) shows the kinetic traces together with the fit at three fluences 0.1, 0.9 and 1.9 mJ/cm$^2$.  Fig \ref{fig:TA_param} (a-c) summarizes the fluence dependence of  filling factor $f_\phi$,  decay time constant obtained from first-order rate constant $k_1^{-1}$ and the constant $A_L$ as a function of pump fluence. From the fit we obtain $k_A=(3\pm 1)\times10^{-28}$ cm$^6$s$^{-1}$.

The filling factor shows a linear decrease with fluence upto 0.6 mJ/cm$^2$, followed by a less slow decrease  for higher fluences (Fig. \ref{fig:TA_param} (a)).   The decay time constant shows a linear increase upto 0.5 mJ/cm$^2$ and saturates to a value of 1.4 ps for higher fluences (Fig. \ref{fig:TA_param} (b)). The linear increase is due to the temperature dependent specific heat which scales as $k_1^{-1}= C_e(T_e)/g$  for low excitation fluences \cite{yu2011excitation,park2007ultrafast}. However, at higher fluences, as the filling fraction decreases, the overall excited carrier density  saturates to its maximal number and the electronic temperature does not increase on further increasing the fluence. This leads to deviation from linear behavior and saturation of $k_1^{-1}$. The fluence dependence of $A_L$ is found to depend sublinearly on the pump fluence (Fig. \ref{fig:TA_param} (c)). As  $A_L$ is proportional to the rise in the lattice temperature $T_L$ its fluence dependence can be understood in terms of lattice heat equation discussed in TTM. Taking the assumption that peak elctronic temperature $T_e >> T_L$ the rise in the lattice temperature is $T_L \propto T_e$ \cite{sun1994femtosecond,sun1993femtosecond}. The peak electronic temperature $T_e \propto \sqrt{\Phi}$, where $\Phi$ is the pump fluence \cite{groeneveld1995femtosecond}. Therefore, rise in lattice temperature $T_L$ and  amplitude $A_L \propto \sqrt{\Phi}$ also. The solid line in Fig. \ref{fig:TA_param} (c) is fit to $A_L$ with $\sqrt{\Phi}$.

\section{\label{sec:level4}  Conclusion}
In conclusion, we have presented a systematic study of nonequilibrium carrier dynamics in Au-UNWs using femtosecond time resolved technique.  The results demonstrate that the carrier dynamics in Au-UNWs of diameter $\sim$ 2 nm can not be fully captured by the well established two-temperature model. The effect of enhanced Coulombic interaction in low-dimensional nanosystems is seen to play a crucial role in carrier dynamics. The reduced screening effect promotes many-body interactions leading to nonequilibrium carrier density dependent Auger heating which slows down the carrier cooling. Our experiments together with the rate equation model gives an estimate of the Auger coefficient $k_A^{-1}$ and the phonon mediated carrier relaxation rate in Au-UNWs. The results are of potential importance for devices and sensors based on Au-UNWs whose performance depends on the electron dynamics in nano-gold \cite{maurer2016templated,gong2014wearable,lu2010nanoelectronics}.    
 
\begin{acknowledgments}
 AKS thanks Department of Science and Technology, India under the Nanomission Project for financial support.  GP thanks Council of Scientific and Industrial Research (CSIR) for SRF. 
\end{acknowledgments}

%

\end{document}